\documentclass[aps,prl,twocolumn]{revtex4}

\begin{document}

\title{Non-cubic layered structure of Ba$_{1-x}$K$_{x}$BiO$_{3}$ superconductor}

\author{L.A. Klinkova$^{1}$, M. Uchida$^{2}$, Y. Matsui$^{2}$,
V.I. Nikolaichik$^{3}$, N.V. Barkovskii$^{1}$}
\address{$^{1}$Institute of Solid State Physics, Chernogolovka, Moscow 
District 142432, Russia \\$^{2}$Advanced Materials Laboratory, National Institute 
for Materials Science, Tsukuba 305-0044, Japan\\$^{3}$Institute of 
Microelectronics Technology, Chernogolovka, Moscow District 142432, Russia}

\begin{abstract}
Bismuthate superconductor Ba$_{1-x}$K$_{x}$BiO$_{3}$ (x=0.27-0.49,
T$_{c}$=25-32K) grown by an electrolysis technique
was studied by electron diffraction and high-resolution electron microscopy.
The crystalline structure thereof has been found to be non-cubic,
non-centrosymmetric and of the layered nature,  with the lattice parameters 
a$\approx$a$_{p}$, c$\approx$2a$_{p}$ (a$_{p}$ is a simple cubic perovskite 
cell parameter) containing an ordered arrangement of barium and potassium.
The evidence for the layered nature of the bismuthate superconductor removes
the principal crystallographic contradiction between bismuthate and cuprate
high-T$_{c}$ superconductors.
\end{abstract}


\maketitle

Superconducting bismuthates including the firstly discovered 
BaPb$_{1-x}$Bi$_{x}$O$_{3}$ [1] and Ba$_{1-x}$K$_{x}$BiO$_{3}$ (BKBO) [2] 
with the highest transition temperature (T$_{c}$=32-35K for x$\approx$0.4) have 
the basic characteristics similar to those of high-T$_{c}$ copper-oxide 
superconductors [3]. This might favor for a common pairing 
mechanism. On the other hand, bismuthates have been generally considered to be 
fundamentally different from cuprates due to two reasons. Bismuthates are 
non-magnetic, and they have three-dimensional structures rather than layered 
two-dimensional ones characteristic of cuprates. Of these two contradictions,
the most important one seems to be the latter. The absence of magnetic fluctuations
in bismuthates may point to a non-magnetic nature of the pairing mechanism, whereas
it is unlikely that a common superconducting scenario in cuprates and bismuthates
does not depend on lattice dimensionality.

The study of BKBO is of a particular interest also for understanding 
the relation between charge-density-waves (CDWs) and superconductivity in
high-T$_{c}$ oxides. The parent compound for BKBO is the perovskite BaBiO$_{3}$
containing a CDW formed of an ordered arrangement of non-equivalent
bismuth ions referred to as Bi$^{3+}$ and Bi$^{5+}$ [4]. This CDW is assumed to
be responsible for the semiconducting behavior of BaBiO$_{3}$ and
Ba$_{1-x}$K$_{x}$BiO$_{3}$ materials with low potassium content (x $<$ 0.25).
The widely known notion that BKBO superconductors have a simple cubic
ABO$_{3}$ solid-solution structure of a non-layered nature, with barium and
potassium randomly occupying the A-position, was inferred from long-range
structural studies of ceramic samples by X-ray [5] and neutron diffraction [6].
A simple cubic structure excludes the existence of a CDW that leads to
the conclusion of the total incompatibility of CDWs and superconductivity.
However, studies of BKBO by methods sensitive to short-range symmetry,
in particular, by Raman scattering spectroscopy [7], a paired-distribution
function analysis of neutron diffraction data [8], extended X-ray fine
structure analysis (EXAFS) [9,10] evidence that the local structure
of BKBO superconductors is not cubic. An X-ray diffraction study [11] of
Ba$_{0.6}$K$_{0.4}$BiO$_{3}$ single crystals grown by an electrolysis 
technique [12] has revealed supercell reflections assigned to a CDW. 

It has been hypothesized [13] that BKBO superconductors may have layered
structures with an anisotropic ordered arrangement of potassium and barium.
Detailed studies [14] of phase relations in the Ba-Bi-O system have revealed
a series of Ba$_{n}$Bi$_{n+m}$O$_{y}$ oxides of ordered layered
structures [15]. Ba$_{n}$Bi$_{n+m}$O$_{y}$ are assumed to be matrices transformed,
when intercalated with potassium, into superconducting oxides retaining the 
layered nature of the matrices structures. Besides, Ba-rich oxides with a
solid-solution structure that attributed to BKBO superconductors were
discovered [16]. Such oxides are formed in two-phase regions of
the Ba-Bi-O system. Assuming that the same phenomenon is inherent also in the
Ba-K-Bi-O system, we looked for conditions of single-phase growth of BKBO
crystals. By varying parameters of the electrolysis technique, we examined the
growth process and obtained the data suggesting the existence of individual
superconducting phases Ba$_{n}$K$_{m}$Bi$_{n+m}$O$_{y}$ with different T$_{c}$
(8-35K) [17].

This paper reports electron diffraction (ED) and high-resolution electron 
microscope (HREM) studies of superconducting crystals Ba$_{1-x}$K$_{x}$BiO$_{3}$ 
(x=0.27-0.49, T$_{c}$=25-32K) grown by a modified electrolysis technique [17].
We present the evidence obtained by a long-range structural method that 
the BKBO superconductor has a non-cubic structure with layered ordering 
of barium and potassium.

Superconducting crystals were produced by electrolysis of 
KOH-Ba(OH)$_{2}$-Bi$_{2}$O$_{3}$ melt (K:Ba:Bi=72:1.33:2) with the current of 5.3
mA at 300$^{o}$C for 5 h. The anode deposit was a polycrystalline boule 
consisting of intergrown single crystals. According to an X-ray powder diffraction
analysis, the deposit contained several phases of pseudocubic 
perovskite structure (a$_{p}$=0.4277-0.4310 nm). The temperature dependence of
the magnetic susceptibility $\chi$ was measured for four crystals of the cubic
shape of 0.5-1 mm$^{3}$ in volume chosen in the deposit. The crystals displayed
similar curves, $\chi$(T), showing bends at 25, 27, 30 and 32K that 
indicates the presence of different superconducting phases.

One of the four crystals was selected for electron microscope studies. It was 
ground to prepare a suspension with particles of a few $\mu$m in size which
was deposited on holey carbon films. Electron diffraction studies were
performed in an electron microscope JEOL JEM-2000FX equipped with a system for 
an energy dispersive X-ray (EDX) elemental analysis. High-resolution studies were 
performed in a microscope Hitachi HF-3000. In order to avoid beam-induced
modulations [18], the experiments were run at minimal electron beam intensity.
We examined 40 particles of the ground crystal, EDX spectra and 
ED patterns were taken from each particle. Although the full composition range
of Ba$_{1-x}$K$_{x}$BiO$_{3}$ particles was found to be x=0.27-0.49, majority
of them had K:Ba:Bi ratio close to average value of 0.38:0.62:1 obtained by
summing over all measurements.

Half of the particles exhibited diffraction patterns (fig.1) containing 
supercell reflections with the vector {\bf q}=$\frac{1}{2}$[001] (indexing in 
terms of a simple cubic perovskite cell) in addition to the basic perovskite
reflections. As only two out of three $\langle$100$\rangle$ axial directions are
observable at electron microscope studies (due to a limited range of specimen
tilt), some of particles not displaying supercell reflections were, in fact,
also of the ordered nature, but those particles were observed in the [001] zone
axis when the supercell reflections could not be excited. Assuming that the 
predominant cleavage at crystal grounding is weak for BKBO superconductors,
a lower limit of the relation of all ordered particles to non-ordered can be
estimated from the above data as 3:1. It indicates that the bulk of the
studied crystal was of the ordered state. From the temperature dependence of
the magnetic susceptibility, which changed much and sharp at 30K and
smoothly at lower temperatures, it follows that the ordered state is related
to T$_{c}$=30K, whereas the non-ordered part of the crystal became
superconducting at T$_{c}$=25-27K.

The anisotropic orientation of the supercell reflections along only one of the
axial directions indicates that the ordered lattice is of non-cubic symmetry.
As no clear splitting of perovskite spots was observed, which would indicate the 
presence of twins of a phase with lower crystallographic symmetry, it follows 
that tetragonal symmetry can be assigned to the ordered lattice. From the 
magnitude and type of the supercell reflections one may state that the supercell
is primitive with the the lattice parameters a$\approx$a$_{p}$ and
c$\approx$2a$_{p}$, it consists of two perovskite blocks.

A [100] HREM image taken from the particle, which displayed supercell 
spots in ED patterns, is shown in fig.2a. The ordering manifests itself as
intensity modulations along the [001] direction with the period of 2a$_{p}$.
A detailed picture of the image contrast is shown in fig.2b presenting the 
enlarged image of the area near the edge of the particle. It can be visualized
that the supercell consists of two perovskite blocks. Matching the corresponding 
spots in the image and measurements of the distances between them disclose that
the image contrast of the blocks is different, and the blocks are slightly different
in size (4-5\%). This indicates the the blocks in the supercell are not equivalent.
An important feature of the HREM image in fig.2b can be noticed, namely, 
the absence of a symmetry center, which suggests a non-centrosymmetric nature of 
the supercell structure.

Two crystallographic models of a different nature may be responsible for the
appearance of supercell reflections with the vector {\bf q}=$\frac{1}{2}$[001]:
(1) the model of solid solution with the occurrence of common barium and potassium
positions in the perovskite blocks, (2) the model of ordering of barium and
potassium.

In the model of solid solution the appearance of the supercell reflections is 
related with distortions (tilting and/or deformation) of oxygen octahedra
surrounding bismuth ions. This approach was applied to explain the origin of
supercell reflections with the vectors
{\bf q}=$\frac{1}{2}$$\langle$111$\rangle$ observed in diffraction patterns
of the perovskite BaBiO$_{3}$ [4]. For our case, a scheme of one-dimensional
deformation of oxygen octahedra along the [001] direction without tilting,
when octahedra are alternatively different in size with the period of 2a$_{p}$,
can only give rise to reflections with {\bf q}=$\frac{1}{2}$[001], because the
a$_{p}$xa$_{p}$x2a$_{p}$ supercell is not compatible with the presence of
tilting [19]. In this scheme, there is a non-equivalency of bismuth ions in
neighbouring planes rectangular to the [001] axis, which is interrelated with
the existence of a CDW.

To establish if the solid solution model is credible, we performed HREM image
simulations using the NCEMSS program [20] with variation of possible parameters
affecting the image contrast: ion coordinates, specimen thickness, objective lens
defocus value, specimen deviation from the exact zone orientation. The
simulation has revealed that no asymmetry in HREM images of centrosymmetric
cells is observed, whereas it does occur for non-centrosymmetric ones. However,
a trustworthy agreement between simulated images of non-centrosymmetric cells
and the experimental images cannot be achieved by varying the parameters. So,
we conclude from structural studies that the model of solid solution should be
rejected. The model of only oxygen octahedra deformations is not compatible also
with superconducting properties of BKBO crystals of the compositions studied here.
Quite intensive reflections with {\bf q}=$\frac{1}{2}$[001]
(more intensive than reflections with {\bf q}=$\frac{1}{2}$$\langle$111$\rangle$ 
characteristic of semiconducting BaBiO$_{3}$) would imply large octahedra deformations
and hence the existence of a very strong CDW, which would result in dielectric properties.

In the model of ordering one of the two cation positions between the oxygen
octahedra (A-positions) is fully occupied by barium. The second A-position is a 
common site for barium and potassium with the highly prevailing amount of 
potassium (for Ba$_{0.6}$K$_{0.4}$BiO$_{3}$ composition being close to 
the average one of investigated particles, the ratio of the occupation 
factors Ba:K is 1:4 in the second A-position). The image simulations (Fig.2c)
have shown that the contrast of experimental images can be described in the
model of ordering with non-symmetric ion coordinates. Intensity of the
supercell reflections is dictated by a difference in scattering factors
of Ba and K.

Fig.3 illustrates the model of the structure consistent with the 
HREM images. The tetragonal space group P4mm can be taken as a possible
non-centrosymmetric group to describe the lattice symmetry. Barium and
potassium are displaced against the centers of the perovskite blocks in 
parallel manner. Apical oxygens O(2) and O(4) are synchronously displaced 
from the centers of the edges, that results in distinct positions thereof
relative to bismuth ions. Bismuth ions in neighbouring BiO$_{2}$ planes
with planar oxygens O(1), O(3) are non-equivalent in this model as well as
in the solid solution one.

The authors of initial studies [21] of BKBO superconductors by EXAFS
assumed the existence of only one Bi-O bond length in the structure.
The recent EXAFS study [10] has found two distances of oxygen relative to
bismuth. It can be assumed that the bond lengths of bismuth with planar
oxygens O(1), O(3) (Fig.3) are close or the same, whereas the bond lengths
of bismuth with apical oxygens are clearly different. As the number of apical
bonds is smaller than that of planar ones, the contribution of the formers
into the general set of Bi-O bonds is insignificant and gives rise to the
second-order effect [10]. However contrary to the assumption put forward in
Ref.10 about the dynamic nature of the occurrence of different Bi-O bonds in
BKBO superconductors, our study shows that it is of a static nature.

BKBO superconductors have the formal oxidation state of bismuth ions in the
range of 4.35-4.55 [17,22]. A part of bismuth ions may be present in the stable
oxidation state +3. It is known that Bi$^{3+}$ ions in the structures
of bismuth-containing oxides are coordinated with surrounding oxygen ions
asymmetrically due to the presence of stereoscopically active lone electron
pairs of Bi$^{3+}$ ions. It can be assumed that a collinear orientation of
these electron pairs gives rise to the asymmetric structure in fig.3.

It is significant that the existence of non-equivalent metal-oxygen distances in 
the ordered structure of BKBO points to the likely presence of a CDW. This CDW
is obviously other than the semiconducting CDW existing in the parent perovskite
BaBiO$_{3}$, because the non-equivalent Bi ions in BaBiO$_{3}$ have all (six)
different Bi-O bonds, whereas such ions in the ordered BKBO differ with respect
to one or two apical oxygens. This is consistent with the X-ray study [11] giving
an evidence that splitting of the supecell reflections resulting from the domain
structure was sensitive to electric and magnetic fields. Further studies with the
use of a quantitative diffraction technique (e.g. neutron diffraction) are needed
to reveal the nature of the CDW and details thereof which may be related to the
pairing mechanism.

An important point is that the ordered arrangement of barium and potassium ions 
along the [001] axis results in an anisotropic structure, where stacking
of BiO$_{2}$ planes with planar oxygens O(1),O(3) along the [001] axis (fig.3)
endows the structure with the layered nature. This makes the crystalline
structure of the ordered BKBO with the separate BiO$_{2}$ planes qualitatively
similar to crystalline structures of cuprate superconductors with CuO$_{2}$ planes
that cancels their principal crystallographic contrasting. The difference between
them is that cuprates have layered structures with distinct anisotropy, whereas
the layered BKBO has a weakly anisotropic structure due to close ion radia of
Ba$^{2+}$ (0.138 nm) and K$^{+}$ (0.133 nm).

In summary, we have observed a non-cubic structure of the superconducting
bismuthate Ba$_{1-x}$K$_{x}$BiO$_{3}$ (x=0.27-0.49, T$_{c}$=25-32K).
The structure is ordered with the lattice parameters a$\approx$a$_{p}$, 
c$\approx$2a$_{p}$ in terms of a simple cubic perovskite cell parameter
a$_{p}$. It has been revealed that the ordering nature is related to the
ordered arrangement of barium and potassium ions, which endows
the structure of Ba$_{1-x}$K$_{x}$BiO$_{3}$ with a layered character being
typical for structures of superconducting cuprates. The crystalline cell of
the superconductor has been found to be non-centrosymmetric with non-equivalent
metal-oxygen distances that may point to the existence of a charge-density wave.

The work at Chernogolovka is supported in part by the Superconductivity 
section of the Russian State Research and Engineering program "Topical Problems 
in the Physics of Condensed Matter" and the project no. 02-02-1678 of the RFBR.

\begin{figure}
\caption{Electron diffraction patterns of ordered particles displaying 
supercell spots (arrowed) with the vector {\bf q}=$\frac{1}{2}$[001]. a) [100] 
zone axis; b) [1-10] zone axis.}
\label{fig1}
\end{figure}

\begin{figure}
\caption{a) A [100] HREM image of a particle with the ordering. b) An enlarged
fragment of the fig.2a image (a dotted rectangle). One may notice
the absence of a symmetry center and slightly different sizes of the perovskite
blocks in the supercell. c) A simulated HREM image of the non-centrosymmetric
supercell (drawn in fig.3) with an ordered arrangement of barium and potassium
ions.}
\label{fig2}
\end{figure}

\begin{figure}
\caption{The [100] projection of a non-centrosymmetric a$_{p}$xa$_{p}$x2a$_{p}$ 
supercell of the Ba$_{1-x}$K$_{x}$BiO$_{3}$ superconductor. The size of the 
perovskite block with potassium is slightly smaller than that of with barium. 
Symmetry planes of the blocks are marked by dotted lines.}
\label{fig3}
\end{figure}

\end{document}